\pdfoutput=1
\documentclass[12pt,preprint]{aastex}

\slugcomment{}

\shorttitle{}
\shortauthors{}

\begin{document}

\title{DISPERSION OF OBSERVED POSITION ANGLES OF SUBMILLIMETER POLARIZATION IN MOLECULAR CLOUDS}

\author{G. Novak,\altaffilmark{1} J. L. Dotson,\altaffilmark{2} and 
H. Li\altaffilmark{3}}

\altaffiltext{1} {Department of Physics and Astronomy, Northwestern
University, Evanston, IL 60208, \mbox{g-novak@northwestern.edu}}

\altaffiltext{2} {NASA/Ames Research Center, MS 245-6, Moffett Field, CA
94035}

\altaffiltext{3} {Harvard-Smithsonian Center for Astrophysics, Cambridge,
MA 02138}

\begin{abstract}

One can estimate the characteristic magnetic field strength in GMCs by comparing submillimeter polarimetric observations of these sources with simulated polarization maps developed using a range of different values
for the assumed field strength.  The point of comparison is the degree of order in the distribution of polarization position angles.  In
a recent paper by H. Li and collaborators, such a comparison was carried out using SPARO observations of two GMCs, and employing simulations by E. Ostriker and collaborators.  Here we reexamine this same question, using the same data set and the same simulations, but using an approach that differs in several respects.  The most important difference is that we incorporate new, higher angular resolution observations for one of the clouds, obtained using the Hertz polarimeter.  We conclude that the agreement between observations and simulations is best when the total magnetic energy (including both uniform and fluctuating field components) is at least as large as the turbulent kinetic energy.  

\end{abstract}


\section{Introduction}

The importance of numerical turbulence simulations for molecular cloud studies is now well established, due 
to increasingly detailed observations that confirm the clouds' turbulent nature (e.g., Brunt \& Heyer 
2002), and due to improvements in the simulations themselves, which can now account for many observed 
properties of molecular clouds (Mac Low \& Klessen 2004, and references therein).  The relative importance 
of turbulence vs.\ strong magnetic fields for controlling the time scale of star formation and for 
determining the masses of stars is now a topic of intense debate (Ballesteros-Paredes et al.\ 2007; Padoan 
\& Nordlund 2002; Girart et al.\ 2006; Shu et al.\ 2004).  A key goal of molecular cloud research is to 
determine the structure and strength of the uniform and fluctuating (i.e., turbulent) components of the 
magnetic field.  The most direct method is to measure the field strength via the Zeeman effect (Crutcher 
1999, 2004).  However, these measurements are difficult and subject to selection effects (Bourke et al.\ 
2001), so it is important to try other approaches as well.  

Chandrasekhar \& Fermi (1953, hereafter CF) obtained a reasonably accurate estimate for the field strength 
in the diffuse ISM by using measurements of the interstellar polarization of background stars together with 
a simple formula which they derived.  The CF formula directly relates the dispersion in polarization position angle 
to the ratio of two energy densities: the energy density of the uniform component of the field and the 
energy density of turbulence.  If the uniform field is stronger (weaker) than turbulence, the angle 
dispersion is low (high).  The polarimetry observations that CF used were taken along lines of sight nearly 
perpendicular to the ambient Galactic field, so they were able to ignore projection effects.  A very promising method for 
probing the magnetic field strength in molecular clouds is to apply the CF technique to 
submillimeter polarimetric observations of these clouds (Crutcher et al.\ 2004).  However, in this case 
one does have to worry about projection effects, since the inclination of the field to the line of sight is 
generally unknown.

Li et al.\ (2006; hereafter Li2006) obtained submillimeter polarimetric maps for four Giant Molecular Clouds (GMCs), using the SPARO instrument at South Pole.  Their maps represent an advance over
previous work with respect to sky coverage; the maps cover sky areas that are comparable to the sizes of the GMCs themselves.  Li2006 chose not to apply the CF formula to their data, but instead they compare 
their maps directly to the simulated GMC polarization maps of Ostriker, Stone, \& Gammie (2001; hereafter OSG2001).  
These synthetic maps were developed using numerical simulations of magnetohydrodynamic (MHD) turbulence, and three 
different values for the strength of the uniform field component were explored.  Via their comparisons of 
observations with simulations, Li2006 argue that the magnetic energy density must be comparable to the kinetic energy density of the turbulence.  Note that the OSG2001 
polarization maps were actually intended to be compared with optical polarization measurements for stars 
viewed through GMCs. However, as shown by Li2006, the cloud/grain model of OSG2001 is very simple and consequently their simulations can serve equally well as simulated 
submillimeter polarization maps.  

In carrying out their comparisons between observed and simulated polarization maps, Li2006 made use of results for just two of their four GMCs.  The other two GMCs were not used because of problems related to distance and evolutionary stage.  The GMCs that were included in their comparisons are NGC 6334 and G~333.6$-$0.2.  Before carrying out the comparisons, they first smoothed the simulations to match the SPARO beam size.  In \S~4 below, we further discuss the restriction to just two GMCs, as well as the issue of smoothing.  

In this paper, we reexamine the problem of comparing SPARO maps with OSG2001 maps as a method to constrain  field strength.  Whereas Li2006 used the ``order parameter'' (\S~4) as the specific point of comparison, here we employ the dispersion of the position angle.  Another difference is that we make use of
new submillimeter polarimetry results for NGC 6334 obtained with the Hertz polarimeter.  The Hertz data are complementary to the SPARO maps in that they cover smaller sky regions but with better angular resolution.  
We describe the SPARO and Hertz data sets in \S~2 and the OSG2001 simulations in \S~3.  In \S~4 we review the methods and results of Li2006, and our new analysis is presented in \S~5 and \S~6.  

\section{Observations}

\subsection{Large-scale Maps from SPARO}

SPARO is a 9-pixel 450 \micron\ polarimeter developed at Northwestern Univerisity (Novak et al.\ 2003, 
Renbarger et al.\ 2004) for use with the Viper telescope at South Pole station.  The \mbox{2 m} diameter 
Viper telescope was operational between 1997 and its decommissioning in 2006.  
As reported by Li2006, SPARO was used in 2003 to make large-scale polarization maps for four 
well-separated GMCs in the Galactic disk: NGC 6334, G~333.6$-$0.2, G~331.5$-$0.1, and the Carina Nebula.  The beam size was 4.0\arcmin\ (FWHM) and the 
pixel-to-pixel separation was 3.3\arcmin.   Each of the four SPARO 2003 GMC maps extends over a sky area 
corresponding to several hundred arcmin$^2$, whereas earlier far-IR/submillimeter 
polarization studies of GMCs had covered much smaller sky areas, typically of order 10 arcmin$^2$ (Hildebrand et 
al.\ 2000; Dotson et al.\ 2000; Greaves et al.\ 2003).  

\subsection{Small-scale Maps from Hertz}

Hertz is a 32-pixel 350 \micron\ polarimeter developed at the University of Chicago (Schleuning et al.\ 
1997; Dowell et al.\ 1998) for use with the \mbox{10 m} Caltech Submillimeter Observatory (CSO) on Mauna 
Kea.  Hertz was used at CSO during 1994-2003 to map magnetic fields in star forming clouds and in the 
Galactic center (Hildebrand et al.\ 2000, and references therein).  Here we present new Hertz results for two intensity peaks in NGC 6334, obtained during February 1998 and January 1999.  These data are also included in an ``archival'' paper that has been submitted for publication (Dotson et al.\ 2008).  In Figure~1, these Hertz results are shown as ``blow-ups'' (expanded 
views) of two small regions within the larger-scale SPARO map. 
The Hertz beam size was 20\arcsec\ (FWHM), 
and the pixel spacing was 18\arcsec.  Each of the 
blow-ups in Figure~1 has an associated key indicating the correspondence between bar length and polarization magnitude.  
Note that a given bar length corresponds to a smaller polarization for Hertz than for SPARO (the 
scales differ by a factor of 1.5).  All of the Hertz measurements shown in Figure~1 have 3$\sigma$ significance. 

\section{Numerical Simulations by OSG2001}

OSG2001 use 3-dimensional time-dependent numerical MHD simulations to follow the non-linear evolution of 
initially smooth, self-gravitating, isothermal gas.  The initial magnetic field is uniform, and the initial 
velocity field corresponds to Kolgomorov-type turbulence with sonic Mach number $M$ $\sim$ 14.  
``Snapshots'' of density, velocity, and magnetic field structure are obtained for $M$ = 9, 7, and 5; i.e. 
after much of the turbulent energy has been allowed to decay.  The intention of OSG2001 is not to provide 
models of the time evolution of GMCs.  Rather, it is the snapshots themselves that are intended to 
represent GMCs, and the time evolution is merely a convenient method for developing them.

OSG2001 parametrize the the energy in the uniform field component using the ``thermal beta'', defined as 
$\beta$=$c_{s}^2/v_{a}^2$, where $c_{s}$ is the sound speed and $v_{a}$ is the Alfven velocity. Three 
values of $\beta$ are used which we will refer to as the strong field ($\beta$ = 0.01), ``intermediate'' 
field ($\beta$ = 0.1), and weak field ($\beta$ = 1) cases.  The authors make the simplifying assumption 
that dust having uniform polarizing efficiency is mixed in with the gas, and on this basis they 
calculate the optical polarization that would be observed toward background stars viewed through the 
simulated GMCs.  These stellar polarization simulations are done for the $M$ = 7 case only.  Each of the 
three field-strength cases (weak, intermediate, and strong), corresponds to a specific value for the 
ratio $R_{U:K}$ of the energy density of the uniform field to the kinetic energy density.  This can be 
found from $R_{U:K} = M^{-2}\beta^{-1}$. The value of the Mach number $M$ actually varies slightly for 
the ``$M$ = 7'' snapshots, averaging around 7.4, so the three field-strength cases correspond 
approximately to $R_{U:K}$ = 0.018 (weak), 0.18 (intermediate), and 1.8 (strong).

OSG2001 present histograms (their Fig.~24) of the distribution of stellar polarization angle for 15 
separate cases, corresponding to all combinations of the three values of $\beta$ with five assumed 
values for the angle $\theta$ by which the mean magnetic field is inclined with respect to the plane of 
the sky ($\theta$ = 0\degr, 30\degr, 45\degr, 60\degr, and 90\degr).  For each of the 15 cases, the 
dispersion $\delta \phi$ of the distribution is given, and for two cases, the the actual simulated 
polarization maps are shown (see their Figs.~22 and 23).  Specifically, maps are shown for the case 
where the field is strong ($\beta$ = 0.01) and lying in the plane of the sky ($\theta$ = 0\degr), and 
the case where the field is weak ($\beta$ = 1) and lying in the plane of the sky.  The authors also made 
a $\theta$ = 0\degr\ map for the intermediate field ($\beta$ = 0.1) case, but they did not publish it.  
However, one of us (H.L.) obtained it directly from E. Ostriker while preparing Li2006. 
OSG2001 make use of their $\delta \phi$ values to test the CF formula (\S~1).  They conclude that the 
formula accurately reproduces the strength of the field component lying in the plane of the sky, 
provided that (a) one restricts to cases where $\delta \phi <$ 25\degr, and (b) one applies a 
multiplicative correction factor.

In Figure~2 we include a synopsis of the 15 values of dispersion $\delta \phi$ that are derived by OSG2001 
from their 15 histograms.  For each of the three values of $\beta$, we show the total range covered by the 
five corresponding values of $\delta \phi$.  Note that we chose $\beta^{-1}$ as the abscissa for the plot, 
so field strength increases to the right.  The weak field and intermediate field cases show high dispersion 
regardless of $\theta$, indicating that the uniform component of the field is dominated by the fluctuating 
field component.  The strong field case, on the other hand, can have very low dispersion; it varies from 
less than 10\degr\ (for $\theta$ = 0\degr) to more than 40\degr\ (for $\theta$ = 90\degr).  Note that the 
highest values of $\delta \phi$ seen in Figure~2 are near 50\degr.  This approximately corresponds
to the dispersion value of $12^{-0.5}\pi$ = 51.96\degr\ that one obtains for the case of a random distribution of 
polarization position angles (Serkowski, 1962).   

For each of the 
three values of $\beta^{-1}$, we also show a solid dot that represents an estimate of the expectation value $\langle \delta 
\phi \rangle$ of the dispersion.  We derived this by averaging together the five values of $\delta \phi$ 
given by OSG2001 for each field-strength case, weighting each by the likelihood of its occurrence, under 
the assumption that all possible values for the orientation of the field in 3-dimensional space are equally 
likely.  Specifically, the expectation values were derived by numerical evaluation of \begin{displaymath} 
\langle \delta \phi \rangle = \int_{0}^{90} \delta \phi (\theta ) \, 2 \pi \cos \theta \, d\theta \, \, \, 
\, \, \, , \end{displaymath} where for every discrete value $\theta_{i}$ used in the numerical integration, 
in place of $\delta \phi (\theta_{i})$ we simply use whichever one of the five values of $\delta \phi$ 
given by OSG2001 ($\delta \phi (\theta = 0\degr)$, $\delta \phi (\theta = 30\degr)$, $\delta \phi (\theta = 
45\degr)$, $\delta \phi (\theta = 60\degr)$, or $\delta \phi (\theta = 90\degr)$) has $\theta$ closest to 
$\theta_{i}$.

\section{Methods and Conclusions of Li2006}

Li2006 draw two main conclusions from their analyses of the SPARO GMC observations discussed above (\S~2.1):  First, they report evidence of continuity between Galactic fields and GMC fields.  This evidence comes from comparisons of GMC field directions with Galactic plane orientation and with stellar polarization data.  We will not discuss these issues here.  Their second conclusion relates to magnetic field strength in GMCs, and follows from comparisons of the observed polarization maps with the simulated maps of OSG2001.  We next review this second conclusion and the methods used to reach it.  

The numerical simulations by OSG2001 include spatial scales spanning 8 octaves (a factor of 
2$^8$ = 256), ranging from scales comparable to GMC sizes down to scales below 0.1 pc.  The 
SPARO observations of Li2006 extend over barely two octaves of spatial frequency, and 
probe spatial scales no smaller than 2 pc even for NGC 6334, the closest of the four SPARO GMCs.  Thus, the SPARO maps are missing significant information about the degree of field disorder on 
small scales.  To overcome this problem, Li2006 smoothed the OSG2001 maps to SPARO's resolution before comparing them with the observations.

Li2006 reviewed the characteristics of their four targets, and concluded that one of them, the Carina Nebula, was hotter and probably much more evolved than 
the others.  The magnetic field geometry in Carina was found to be dominated by the effects of expanding bipolar super-bubbles; the cloud appears to have been literally torn apart by energy released due to formation of massive stars in the recent past.  Li2006 conclude that it would make little sense to compare the bubble-like geometry of Carina's magnetic field with the turbulent cascade simulations of OSG2001.  Unlike Carina, the other three clouds (NGC 6334, G~333.6$-$0.2, and G~331.5$-$0.1) appear to 
be typical high mass GMCs with evidence for ongoing formation of high mass stars (Li2006). 

As explained in \S~3, Li2006 had access to OSG2001 polarization maps corresponding to all three values of $\beta^{-1}$, but for the $\theta$ = 0\degr\ case only (field parallel to sky plane).  The OSG2001
maps correspond to 8 pc regions (see Li2006).  Each of the three simulated polarization maps was smoothed by Li2006 to 2 pc resolution (to make three simulated NGC 6334 maps) and 4 pc resolution (yielding three simulated G~333.6$-$0.2 maps).  The three simulated NGC 6334 maps contain 16 vectors each, while the three simulated G~333.6$-$0.2 maps contain only four vectors 
each.  
It is impossible to degrade the resolution of the simulations sufficiently to allow for comparisons with G~331.5$-$0.1, the most
distant of the four GMCs.

In order to compare the degree of magnetic field disorder in 
the SPARO observations of NGC 6334 and G~333.6$-$0.2 with that seen in the corresponding smoothed simulations, Li2006 calculate an ``order parameter'' for each map.  This parameter varies between 0 (total disorder) and 1 (uniform field).  They find that the smoothed simulated maps corresponding to weak and intermediate field cases give 
order parameters that are too low to be consistent with the corresponding SPARO maps, while the strong field maps yield order parameters that are too high.  

However, Li2006 point out that the histograms given in Figure 24 of OSG2001 indicate that the degree of order in the strong field simulated map drops as $\theta$\ increases (as discussed in \S~3).  Non-zero values of $\theta$\ would therefore result in lower values for the order parameter, reducing the discrepancy between the strong field case and the observations.  On the basis of this qualitative argument, they favor the strong field model ($\beta^{-1}$ = 100) as the best match to their data.

\section{Dispersion Estimates for NGC 6334 and G333.6$-$0.2}

A key difference between the analysis we present here and that of Li2006 is that we use dispersion in polarization position angle to quantify the degree of disorder, for both the SPARO observations and simulated GMC maps, whereas Li2006 used the order parameter.  Another difference is that instead of smoothing the simulations to match the resolution of the observations, we calculate dispersion values directly from the SPARO data and then we apply upward corrections to account for the small-scale disorder in polarization angle to which the SPARO observations are 
insensitive.  The goal of our corrections is to account for all ``relevant'' small-scale 
disorder, i.e., extending down to the smallest scales sampled by the simulations of OSG2001.  We will 
refer to the resulting corrected dispersion estimates as the ``8-octave-corrected'' dispersion estimates because 
they are intended to include the effects of disorder over eight octaves in spatial frequency, just as 
the simulations do (\S~4).  We derive the 8-octave-corrected dispersion estimates for both NGC 6334 and G~333.6$-$0.2 in this section.  Subsequently, in \S~6, we compare these values to the dispersion values derived from the OSG2001 models.

We employ two different methods to obtain the above-mentioned corrections.  ``Method 1'' is based on the OSG2001 models:  We use these to estimate the factor by which the dispersion is reduced when smoothing the simulated polarization maps to 
the 2 pc (4 pc) resolution of the SPARO NGC 6334 (G~333.6$-$0.2) observations.   For each of the two clouds, we then multiply the dispersion calculated from the SPARO maps by the reciprocal of this reduction factor.
The smoothed maps that we will use for these calculations are identical to
those employed by Li2006 (see \S~4). Our second method for deriving the needed corrections (``Method 2'') works only for NGC 6334, and involves using 
the Hertz data (\S~2.2 and Fig.~1) to obtain information about small-scale disorder.  Our analysis has advantages as well as disadvantages, in comparison with that given by Li2006.  We discuss these in \S~6.

The starting point is to calculate the dispersion in polarization position angle for the 
SPARO maps of NGC 6334 and G~333.6$-$0.2, using values of $\phi$ given in Table 1 and Table 
3\footnote{We discovered a transcription error in Table 3 of Li2006.  The value of $\phi$ for 
the (0,0) position should be 149.9\degr, not 59.9\degr.  One of us (H.L.) is preparing an erratum 
report.} of Li2006.  We correct for the the variation in 
$\phi$ due to measurement uncertainty, as follows:  For each cloud, we compute
the root mean square of the tabulated $\sigma_{\phi}$ values and then
carry out a quadrature subtraction of this quantity from the nominal dispersion derived from 
the tabulated values of $\phi$.   The results are 22.3\degr\ and 21.6\degr\ (see Table 1).  The measurement uncertainty corrections have only a small effect ($\sim$ 
1\degr). 

The remainder of this section deals with estimating the corrections that must be applied to 
these SPARO dispersion values in order to obtain the desired 8-octave-corrected dispersion estimates for our two clouds.  We first consider NGC 6334.  We begin with Method 1, which requires comparing dispersion values
for the OSG2001 maps to those calculated for our smoothed versions of these maps.  Recall that OSG2001 maps are available for
all three field strength cases, but only for the $\theta$ = 0\degr\ case (\S~3).  Also recall that the dispersion values given by OSG2001 for the weak and intermediate field cases approach the $\sim$ 50\degr\ value corresponding to a random distribution (\S~3; Fig.~2).  Our smoothed versions of the weak and intermediate field maps have a highly disordered appearance, suggesting similarly high dispersion values.  Calculating dispersion for such highly disordered maps is non-trivial, since the mean polarization angle is undefined (Li2006).  
OSG2001 solved this problem by fitting Gaussians to the distributions.  In this paper we adopt a different solution, which is to calculate
the dispersion relative to the equal weight Stokes mean (EWSM) polarization angle (Li2006).  
For the unsmoothed 
intermediate field map, we calculate a dispersion of 49.3\degr\ using our technique.  This is very similar to the dispersion value derived by OSG2001 using their method, which is 50.8\degr.   When the intermediate field map is smoothed to 2 pc, we calculate a dispersion of 52.1\degr, and smoothing to 4 pc resolution yields 50.3\degr.  We conclude that for these highly disordered maps, smoothing has almost no effect on the dispersion.  This is not surprising, since they
are very disordered on all scales and dispersion values approach the maximum.  Results for the
weak field case are similar.  

For the strong field map, the dispersion before smoothing is 9.3\degr.  (This is the smallest $\delta \phi$ value in Fig.~2, located at the bottom end of the range given for $\beta^{-1}$ = 100.)  Smoothing to the 2 pc resolution of our NGC 6334 SPARO map reduces the dispersion to 6.8\degr, corresponding to a factor of 1.37 decrease.  The desired correction factor for NGC 6334 could thus lie anywhere between $\sim$1.0 (weak and intermediate field cases) and 1.37 (strong field case).  
Also note that if we had access to strong field simulated polarization maps for non-zero values of $\theta$, we would likely find that the correction factor would drop to $\sim$1.0 for $\theta$ approaching 90\degr, since the $\theta$ = 90\degr\ strong field polarization map has high dispersion (\S~3).  Since we don't know $\beta^{-1}$ and $\theta$ for the SPARO GMCs, there is ambiguity in the correction factor.  
For the purpose of making a rough estimate, however, we take the geometric mean of the two extreme possibilities for the correction factor to obtain (1.37 $\times$1.0)$^{0.5}$ = 1.17.  We multiply the dispersion of the NGC 6334 SPARO data by this factor to obtain a very rough estimate for the 8-octave-corrected dispersion for the NGC 6334 cloud.  The result is 26.1\degr\ (Table 1).

We now estimate the 8-octave corrected dispersion for the NGC 6334 cloud using Method 2, which makes use of the Hertz observations (Fig.~1, \S~2.2).
The Hertz data are complementary to the SPARO data for NGC 6334, in that they 
sample spatial frequencies well below the SPARO limit of 2 pc, but with very little gap in spatial frequency 
coverage between the two sets of observations.  This can be seen by noting that (a) the diameter of each of 
the circular ``blow-ups'' of Figure~1 was chosen to be equal to the diameter of SPARO's beam (4\arcmin), and (b) 
the Hertz polarization measurements fill a significant fraction of the area within these circles.  Specifically, about half of the combined area of the two circles is
filled in with Hertz vectors.

After correcting for measurement 
error, just as we did for the SPARO data earlier in this section, the northern and southern Hertz fields (Fig.~1) 
yield dispersions of 17.6\degr\ and 17.3\degr, respectively.  The measurement error corrections are again 
small ($\sim$ 1\degr).  Under the assumption that the Hertz results for the two fields are 
representative of the region, we adopt 17.5\degr\ as the dispersion corresponding to spatial scales in 
the range 20\arcsec\ - 3\arcmin, where the lower limit on the range is chosen to be equal to the Hertz 
beam size, and the upper limit is set to the maximal spatial extent of the Hertz maps.  Note that SPARO's spatial frequency coverage corresponds to scales ranging from 4\arcmin\ (SPARO's beam size) to 20\arcmin\ (the spatial coverage of 
the SPARO NGC 6334 map; Fig.~1).  By adding in quadrature the SPARO dispersion (22.3\degr; Table 1) and the Hertz 
dispersion (17.5\degr), we obtain a total dispersion of 28.3\degr\ that is referenced to a total spatial 
frequency coverage of almost six octaves (20\arcsec\ - 20\arcmin).

Six octaves of spatial frequency coverage is close to what we require for comparison with the 
simulations, but we must attempt to estimate the further increase in dispersion associated with 
extending to even higher spatial frequencies, by another factor of four.  No observational data is 
available, but by comparing the SPARO and Hertz dispersion estimates of 22.3\degr\ and 17.5\degr, 
respectively, we can get a rough idea of the shape of the dispersion power spectrum, which we can then 
extrapolate to spatial frequencies above the Hertz range.  Note that, despite the fact that Hertz 
samples a larger range of spatial frequencies than SPARO (3.17 vs.\ 2.32 octaves), Hertz sees a smaller 
dispersion than SPARO.  This suggests that the power spectrum of the dispersion, when expressed in 
square degrees per octave, is falling towards high spatial frequencies.  Specifically, SPARO sees 
(22.3\degr)$^2$ $\div$ 2.32 octaves $=$ 214 sq.\ deg.\ per octave, while Hertz obtains (17.5\degr)$^2$ 
$\div$ 3.17 octaves $=$ 96.6 sq.\ deg.\ per octave.  Via a power law fit to these two points on the 
power spectrum, we can estimate the dispersion corresponding to the $\sim$ two octaves of spatial 
frequency that are too high to be sampled even by Hertz.

To obtain eight octaves of coverage for NGC 6334, we would need to sample scales down to 4.7\arcsec, and 
the missing range (4.7\arcsec\ - 20\arcsec) spans 2.09 octaves.  The geometric mean of the end-points of 
SPARO's spatial frequency range corresponds to 537\arcsec, while that for Hertz is 60\arcsec, and that of 
the ``missing'' band is 9.7\arcsec.  Using these rough values (537\arcsec\ and 60\arcsec) to represent the 
characteristic spatial scales corresponding to SPARO and Hertz data, respectively, and using 9.7\arcsec\ to 
represent the characteristic scale of the missing band, a power law extrapolation yields 49.9 sq.\ deg.\ 
per octave for the missing band at 9.7\arcsec.  The missing band thus has an extrapolated dispersion of 
(49.9 sq. deg. per octave $\times$ 2.09 octaves)$^{0.5}$ $=$ 10.2\degr.  Quadrature addition of this 
dispersion value to the 28.3\degr\ estimate for the dispersion on combined SPARO/Hertz scales yields an 
estimate of 30.1\degr\ for the 8-octave-corrected dispersion.

Our application of Method 2 is only approximate since we have only two observation fields for estimating ``Hertz-scale'' 
dispersion, and since we had to extrapolate to recover the dispersion on even smaller scales.  Method 1 is also very approximate, as noted above.  However, it
is encouraging that the values for the 8-octave-corrected dispersion of NGC 6334 derived via these two very different methods differ by only $\sim$15\% (Table 1).  As indicated in the table, we adopt the average of the two, 28.1\degr, as our best estimate.

Turning to G~333.6$-$0.2, we have no small scale submillimeter polarization data so we estimate the 
corrected dispersion using Method 1 only.  Smoothing the strong field map of OSG2001 to 
4 pc resolution reduces the dispersion from 9.3\degr\ to 5.9\degr, representing a reduction by a factor of 
1.58.  For this cloud, the 
smoothed map contains only four data points (\S~4), so the calculated dispersion is less 
reliable than for NGC 6334.  Nevertheless, just as for NGC 6334, we take the geometric mean of this strong field reduction factor and the value of unity that we 
use as a reduction factor for the intermediate and weak field cases, to obtain a correction factor of 1.26.  This results in an
yields an 8-octave-corrected dispersion of 27.2\degr\ for G~333.6$-$0.2 (Table 1). 

\section{Comparison with Dispersion in Simulated Maps}

Our sample of two clouds is a small one, and the correction for small-scale disorder is uncertain, especially for G~333.6$-$0.2.  Nevertheless, because the SPARO maps are the first submillimeter polarization images to 
sample the structure of the large-scale, or global fields of GMCs, we will compare our best estimates for the corrected cloud dispersions (Table 1) with the $\delta \phi$ values from OSG2001.  This comparison is shown in Figure 2.  It is apparent that the weak and intermediate field 
simulations do not agree well with the observations.  The higher of our two corrected dispersion 
estimates for NGC 6334 (Table 1) is actually marginally consistent with the low end of the 
range of $\delta \phi$ for the intermediate field case, but, overall, the results 
indicate that $\beta^{-1}$ $>$ 10.  

Turning to the strong field case, the range of $\delta \phi$ values given by OSG2001 does overlap with the 8-octave-corrected dispersion estimates that we found for the SPARO clouds (Fig.~2).  However, only one of the five strong field $\delta \phi$ values (corresponding to $\theta$ = 90\degr) is larger than the values from the SPARO clouds.  The second highest is $\delta \phi$ = 
23.7\degr\ which occurs for $\theta$ = 60\degr.  
Thus, the corrected dispersion estimates for the 
SPARO clouds can be brought into agreement with the strong field model only by assuming high 
inclination angles, i.e.\ the field must point nearly along the line of sight ($\theta$ $>$ 60\degr) for both clouds.  If one assumes that the field is a priori equally likely to point in any direction in 3-dimensional space, then
the probability for obtaining $\theta$ $>$ 60\degr\ for two well separated clouds equals (1 $-$ sin 60\degr)$^2$ $\approx$ 1.8\%.  From these arguments, it would appear unlikely that
the strong field model could explain the corrected dispersion values shown in Figure~2, which would lead us to conclude that $\beta^{-1}$ $<$ 100.  However, we will revisit this upper limit later in this section.

Next we summarize the advantages and disadvantages our treatment in comparison to that of Li2006 (\S~4).  
As noted in \S~5, smoothing reduces the dispersion for the strong field map by factors in the range 1.37-1.58,
but has almost no effect on the weak and intermediate field maps.  Our Method 1 correction factor for each cloud is determined by averaging the weak/intermediate field reduction factor and the strong field reduction factor (\S~5).  Thus, these correction factors are quite uncertain.  Because Li2006 directly compare the various smoothed maps to the SPARO data, their method does not suffer from 
this uncertainty, which is an advantage for their treatment.  Our treatment also has advantages.  First, since we use the corrected dispersion as our measure of magnetic disorder, we can
make magnetic disorder comparisons between our clouds and each of the 15 OSG2001 simulations (Fig.~2).  Li2006 are only able to compare with
the three $\theta$ = 0\degr\ maps.  Another advantage of our treatment is that we incorporate information from the Hertz observations.  

As noted above, our use of a single, average Method 1 correction factor for each cloud introduces significant uncertainly into the analysis, since this correction factor actually depends strongly on $\beta^{-1}$ and (for the strong field model) on $\theta$ (see \S~5).  We now discuss the effect of this uncertainty on the reliability of our conclusions.  Starting with our lower limit on $\beta^{-1}$, consider the question of whether the use of erroneous Method 1 correction factors could be the cause of the discrepancy between the high dispersion values for the weak and intermediate field models and the lower values determined for the two SPARO GMCs.  If the true 
$\beta^{-1}$ were actually less than 10, then the Method 1 correction factors we used would be overestimates, by factors of $\sim$1.17-1.26 (Table 1).  If we were to rectify this error by not using the correction factor, however, we would then reduce the corrected cloud dispersion values and thus exacerbate the discrepancy between the weak/intermediate models and the clouds.  Thus, the discrepancy cannot be due to erroneous correction factors, and we conclude that this source of uncertainty does not affect our lower limit on $\beta^{-1}$.  
The same argument cannot be used for our upper limit on $\beta^{-1}$, because in this case the Method 1 corection factor would be expected to vary significantly with incination angle $\theta$ (\S~5).   Thus, the uncertainty in the Method 1 correction factors undermines the arguments we used above to rule out the strong field model.  

In conclusion, our comparison of cloud dispersion values with model dispersion values indicates that field strengths corresponding to $\beta^{-1}$ $>$ 10 are in better agreement with the observational data.  This is consistent with the conclusion reached by Li2006 using a different analysis method (\S~4).  As discussed in \S~3, each value of $\beta^{-1}$ corresponds to a specific value of $R_{U:K}$, the ratio of the energy density of the uniform field to the kinetic energy density.  These correspondences are listed in Table~2.  Our limit on $\beta^{-1}$ gives $R_{U:K}$ $>$ 0.18.  
Using additional information from OSG2001, we can express this constraint in two other ways.  First, OSG2001 specify the ratio 
$R_{F:U}$ of fluctuating to uniform field amplitudes for each of their models.  These $R_{F:U}$ values are also listed in Table~2.   We conclude that $R_{F:U}$ $<$ 2.0.  
Secondly, it is easy to show that the ratio of total field energy (fluctuating plus uniform) to 
turbulent kinetic energy is $R_{T:K}$ = $R_{U:K}$ $\times$ (1 + $R_{F:U}^2$).  The resulting $R_{T:K}$ values comprise the final row of Table 2.  It can be seen that both $R_{U:K}$ and $R_{T:K}$ fall with decreasing $\beta^{-1}$, although $R_{T:K}$ does not fall as rapidly as $R_{U:K}$ because turbulence can pump energy into the fluctuating component of the field.  Our limit on $\beta^{-1}$ corresponds to $R_{T:K}$ $>$ 0.9.

\section{Discussion}

Crutcher (1999) summarized the 27 sensitive Zeeman measurements of magnetic field strengths in molecular 
clouds that were available at that time, and concluded that $R_{U:K}$ $\sim$ 1.  Subsequently, Crutcher et 
al.\ (2004) applied the CF method (with the OSG2001 correction factor; see \S~3) to their submillimeter polarimetric 
measurements of molecular cloud cores and derived field strengths roughly consistent with those found using 
the Zeeman technique (e.g., see comparison in Fig.~3 of Crutcher 2004).  The analysis we have presented 
here is consistent with $R_{U:K}$ $\sim$ 1, though it does not rule out values as low as $R_{U:K}$ $\sim$ 
0.2.  Myers \& Goodman (1991) analyzed optical polarization data for 26 clouds, mostly dark clouds and dark 
cloud complexes, and measured dispersions in the range of 10\degr\ to 25\degr.  Using their own model for
magnetic field structure they concluded that, on average, the fluctuating and uniform field components are approximately equal.  Our limit, $R_{F:U}$ $<$ 2.0, is consistent with this conclusion.

Padoan et al.\ (2004) used numerical MHD turbulence studies to develop simulated molecular clouds.  Their 
work is similar to that of OSG2001 in using an isothermal equation of state, having initially uniform 
magnetic field and density, and having the turbulence driven artificially.  Some differences are that 
Padoan et al.\ (2004) drive the turbulence continuously at $M$ = 10, they do not use self-gravity, and they 
cover a slightly larger range of spatial frequencies ($\sim$8.5 octaves).  Padoan et al.\ (2004) considered 
two field-strength cases, corresponding to $R_{T:K}$ $\sim$ 1 (``equipartition '' case) and $R_{T:K}$ 
$\sim$ 0.2 (``super-Alfvenic'' case).  They showed that the column density structure corresponding to the 
super-Alfvenic case exhibits a power spectrum that is consistent with observations, but that the power 
spectrum for the equipartition case is badly discrepant with observations and is ruled out at the 99\% 
confidence level.  They conclude that magnetic fields in molecular clouds must be fairly weak.  However, our analysis gives $R_{T:K}$ $>$ 0.9, implying fields that are comparable to or stronger than their equipartition case, and much stronger than 
their super-Alfvenic case, which is inconsistent with 
their conclusions.

As noted above, our main conclusions concerning GMC magnetic fields had already been reached by 
investigators working on optical polarimetry (Myers \& Goodman 1991) and on submillimeter polarimetry of 
dense cloud cores (Crutcher et al.\ 2004).  However, our work on the interpretation of large-scale 
submillimeter polarization maps has some advantages over these earlier studies.  Regarding the optical 
polarimetry, there are questions about whether the measurements probe deep into molecular clouds (Arce et 
al.\ 1998, Whittet et al.\ 2001).  Submillimeter polarimetry may be a better probe of molecular regions 
that are well shielded from the interstellar radiation field (Cho \& Lazarian 2005).  Regarding the 
polarimetry of dense cores by Crutcher et al.\ (2004), note that they make use of results from OSG2001, just as we do.  But since our observations
better correspond to the spatial scale of the simulations, we may be on firmer ground than they are.

One limitation of our work on comparing observations and simulations is the overly simple grain alignment 
model used by OSG2001.  There is evidence that even submillimeter polarimetry fails to trace fields when 
one reaches very high densities.  Note, for example, the polarization minima that occur at the peaks of some dense cores (Matthews \& Wilson 2000; Crutcher et al.\ 2004).  
To overcome this problem it will be important to develop turbulence simulations that explore the effects of 
failing grain alignment at high densities (e.g., Padoan et al.\ 2001; Pelkonin et al.\ 2007) but that at 
the same time provide statistics on polarization angle for different assumed values of field strength and 
inclination angle, as in OSG2001.  However, because the coverage of the SPARO maps extends well beyond the 
dense cloud cores and into regions of relatively lower column density, the oversimplified grain alignment 
model may not be as much of a limitation for our study as it is for investigators using CF techniques to 
study dense cores (e.g., Crutcher et al.\ 2004).

Finally, we note that if the polarization angle dispersion in molecular clouds is 
indeed greater than 25\degr, as we have argued, then CF field strength estimates may be of limited value (see discussion in \S~3).  Detailed 
comparison with turbulence simulations using a variety of statistical parameters may be a better way to interpret submillimeter polarization maps of molecular clouds.  

\section{Summary}

We considered the dispersion in polarization angle seen in SPARO maps for two GMCs: NGC 6334 and G~333.6$-$0.2.  
We applied two methods for correcting these dispersion values for the effects of beam 
dilution so that they can be compared with values for simulated GMCs.  Method 1 is to use the simulations themselves to determine the 
necessary correction factor.  Method 2, which is applicable to NGC 6334 only, involves using higher resolution data collected with Hertz as a gauge of small-scale disorder.  The methods are very approximate, but the results are reasonably self-consistent.  When we compare our corrected dispersion values with model dispersion values, we find best agreement for total magnetic energy 
(uniform plus fluctuating components) comparable to or larger than turbulent kinetic energy.

\acknowledgments

This work has benefited from illuminating discussions with R.\ Hildebrand and A.\ Lazarian.  We are 
also grateful to E.\ Ostriker for providing the unpublished simulation, to P.\ Calisse for operating 
SPARO, and to the N.S.F. for supporting SPARO (Award OPP-01-30389 to 
Northwestern U.), Hertz (Award AST-02-04886 to the U.\ of Chicago) and the Caltech Submillimeter 
Observatory (Award AST-05-40882 to Caltech).

\clearpage

\begin{deluxetable}{c c r c}
%\tabletypesize{\scriptsize}
\tablecaption{Dispersion Estimates\tablenotemark{a} for SPARO GMCs}
\tablewidth{0pt}
\tablehead{
\colhead{cloud} &
\colhead{correction method} &
\colhead{dispersion} &
\colhead{notes} 
}

%\colhead{$\Delta\delta$\tablenotemark{a}} &
%\colhead{$P(\%)$} &
%\colhead{$\sigma_{P}$} &
%\colhead{$\phi$\tablenotemark{b}} &
%\colhead{$\sigma_{\phi}$}

\startdata

 NGC 6334  & corrected for $\sigma_{\phi}$ only & 22.3\degr\  & from data in Li2006    \\
 NGC 6334  & Method 1  & 22.3\degr\ $\times$ 1.17 = 26.1\degr\  & 8-octave-corrected    \\
 NGC 6334  & Method 2  & 28.3\degr\  & 6-octave-corrected    \\
 NGC 6334  & Method 2  & 30.1\degr\  & 8-octave-corrected    \\
 NGC 6334  & adopted value  & $\frac{1}{2}$(26.1\degr\ + 30.1\degr) = \bf{28.1\degr}  &   8-octave-corrected  \\
 \hline
 G~333.6$-$0.2  & corrected for $\sigma_{\phi}$ only  & 21.6\degr\  & from data in Li2006    \\
 G~333.6$-$0.2  & Method 1  & 21.6\degr\ $\times$ 1.26 = \bf{27.2\degr} & 8-octave-corrected    \\

\hline
\enddata

\tablenotetext{a}{see \S~5}

%\tablenotetext{b}{$\phi$ is the angle of the E-vector of the polarized
%radiation, measured in degrees from north-south, increasing
%counterclockwise.}

\end{deluxetable}

\clearpage

%\begin{deluxetable}{c r@{.}l r@{.}l r@{.}l}
\begin{deluxetable}{c c c c}

%\tabletypesize{\scriptsize}
\tablecaption{Parameter Values for Model Clouds\tablenotemark{a}}
\tablewidth{0pt}
\tablehead{
\colhead{parameter} &
\colhead{weak field} &
\colhead{intermediate field} &
\colhead{strong field} 
}

%\colhead{$\Delta\delta$\tablenotemark{a}} &
%\colhead{$P(\%)$} &
%\colhead{$\sigma_{P}$} &
%\colhead{$\phi$\tablenotemark{b}} &
%\colhead{$\sigma_{\phi}$}

\startdata

$\beta^{-1}$ & 1 & 10 & 100    \\
$R_{U:K}$ & 0.018 & 0.18 & 1.8    \\
$R_{F:U}$ & 3.5  & 2.0  & 0.52  \\
$R_{T:K}$ & 0.24  & 0.90   & 2.3    \\

 \hline

\hline
\enddata

\tablenotetext{a}{from model of OSG2001; see \S~3 and \S~6}

%\tablenotetext{b}{$\phi$ is the angle of the E-vector of the polarized
%radiation, measured in degrees from north-south, increasing
%counterclockwise.}

\end{deluxetable}

\clearpage


\begin{figure}
\epsscale{0.9}
\plotone{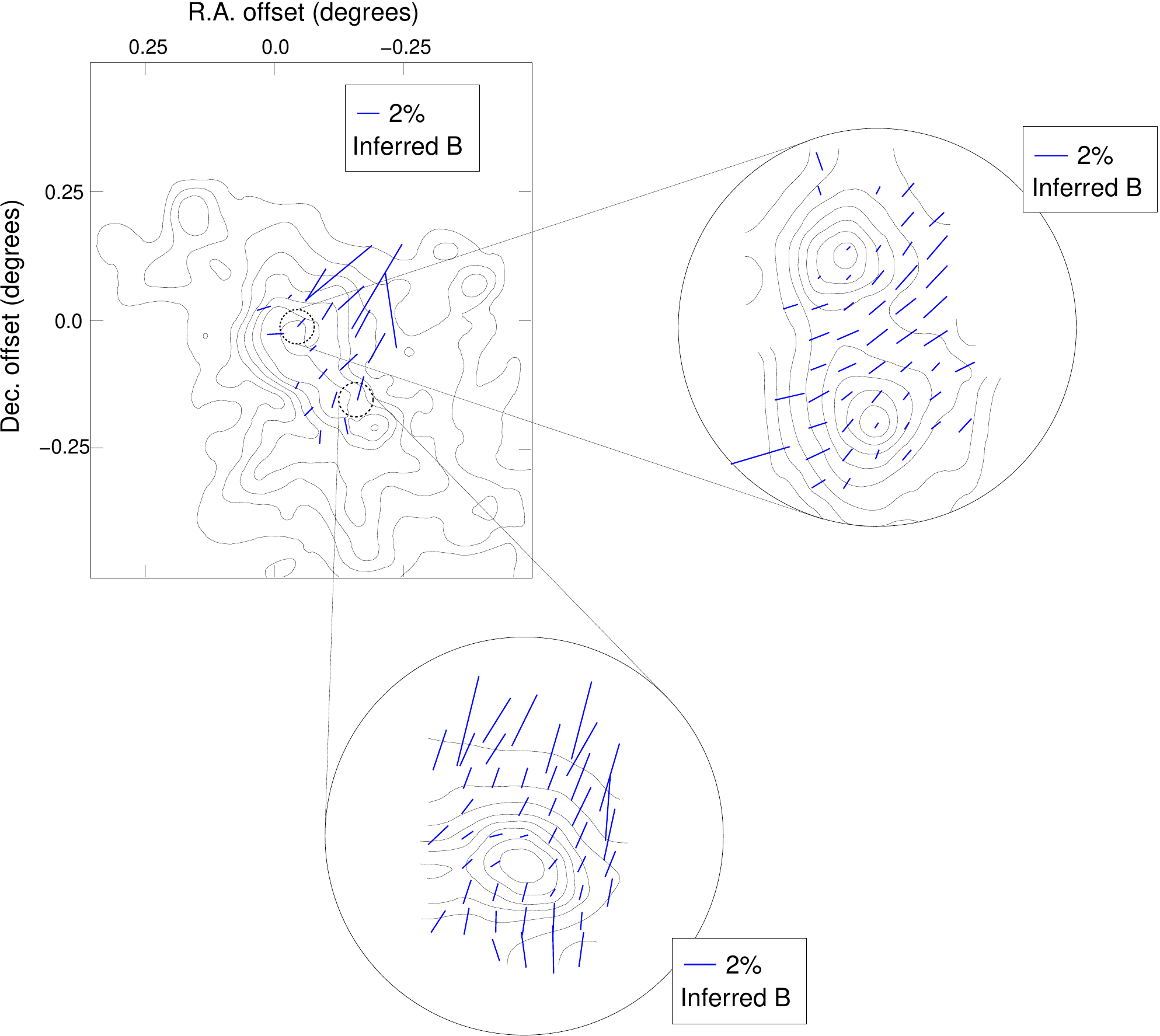}

\caption{ Submillimeter polarimetry of the Giant Molecular Cloud NGC 6334.  The map at upper left (from 
Li2006) shows 450 \micron\ polarization measurements obtained using SPARO, superposed on 100 
\micron\ intensity contours from IRAS ISSA.  For two smaller regions within this map (dashed circles) 
we show expanded views at right and bottom.  These ``blow-ups'' show new 350 
\micron\ polarization measurements obtained using the Hertz instrument.  The contours in the blow-ups 
show the 350 \micron\ total intensity, as measured by Hertz.  For both SPARO and Hertz results, the 
direction of each bar indicates the inferred magnetic field direction (which is perpendicular to the 
measured polarization angle) and the length of the bar is proportional to the degree of polarization.  
Each figure has a key showing the bar length corresponding to a polarization magnitude of 2.0\%.  The 
SPARO map is referenced to the following equatorial (J2000) coordinates: 
($17^h20^m51.0^s$,$-35$\degr$45$\arcmin$26$\arcsec).  Each of the two Hertz intensity maps has contours 
drawn at 20\%, 30\%, 40\%, ..., 90\% of the respective peak flux. }

\end{figure}

\begin{figure}
\epsscale{0.9}
\plotone{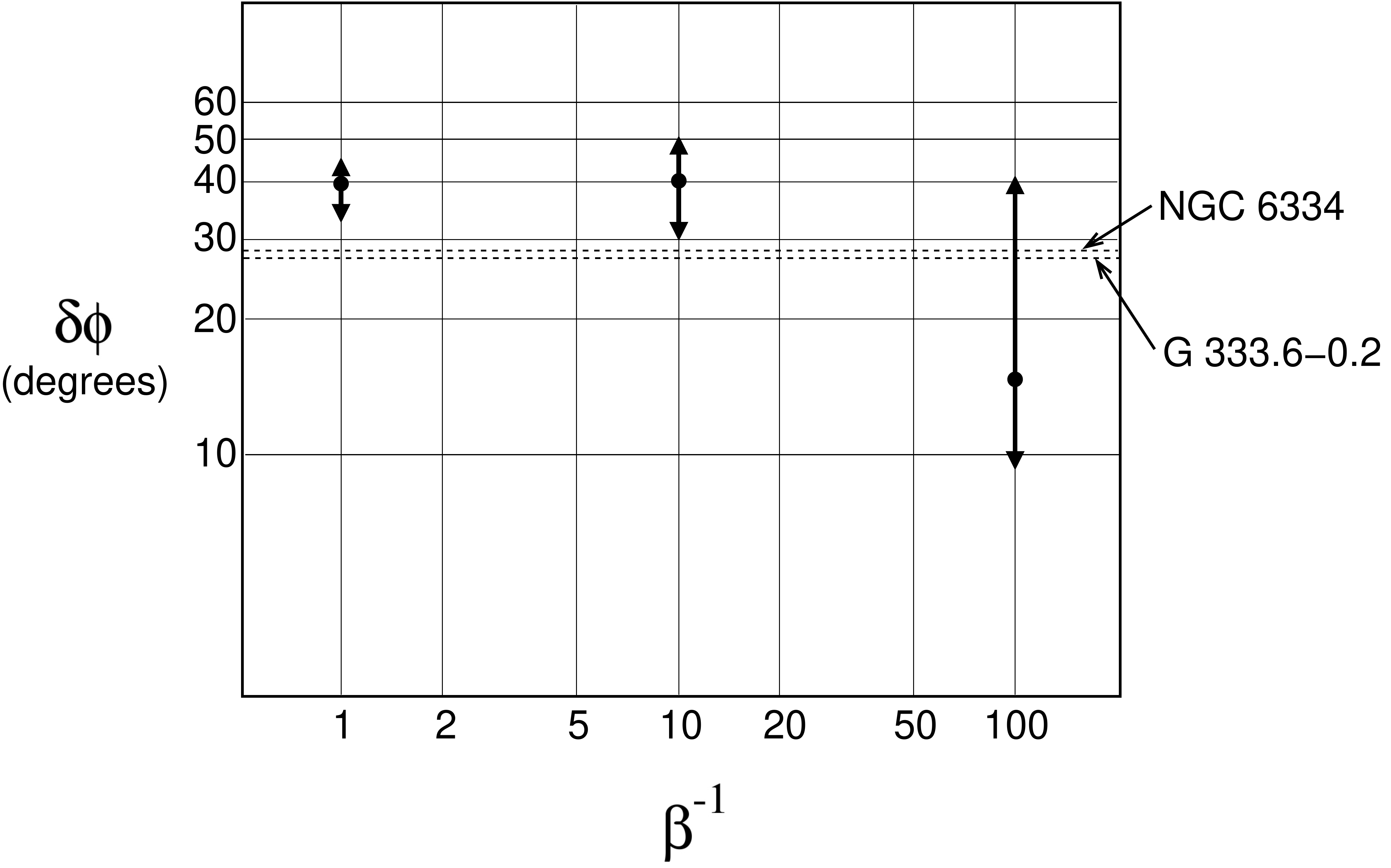}

\caption{ Dispersion $\delta \phi$ in polarization angle vs.\ energy of uniform magnetic field component 
for simulated GMCs (vertical solid lines with arrows) shown together with dispersion estimates based on 
SPARO observations of two GMCs (horizontal dashed lines).  The GMC simulations are from OSG2001 (\S~3).  The energy density of the uniform component of the field is parametrized by the 
inverse of $\beta$.  Simulations were carried out for three values of this parameter, corresponding to 
(from left to right) weak, ``intermediate'', and strong field cases.  For each of the three values of 
$\beta$, five values of $\delta \phi$ were determined, corresponding to five values of the angle $\theta$ 
between the direction of the uniform field and the plane of the sky.  The vertical lines ending in arrows 
show, for each $\beta$, the range covered by the corresponding five values of $\delta \phi$, while the 
superposed dots show the expectation value of $\delta \phi$ (\S~3).  The dispersion 
estimates for the two GMCs include upward corrections to account for small-scale disorder invisible to 
SPARO (\S~5). } \end{figure}

\end{document}